\documentclass[aip,reprint,twocolumn,groupedaddress]{revtex4-1}
\usepackage{color}
\usepackage{graphicx}
\usepackage{amsmath}
\usepackage{latexsym}
\usepackage{graphicx}
\usepackage{grffile}
\usepackage{amssymb}
\usepackage{amsfonts}
\usepackage{amsmath}
\usepackage{mathtools}

\DeclarePairedDelimiter\floor{\lfloor}{\rfloor}
\newcommand{\argmin}{\arg\!\min}

\begin{document}
\title{Parameter estimation, nonlinearity and Occam's razor}

\author{Leandro M. Alonso}
\email{lalonso@rockefeller.edu}
\date{\today}
\affiliation{Laboratory of Mathematical Physics, Center for Studies in Physics and Biology \\ The Rockefeller University, New York, NY 10065, USA.}

\begin{abstract}
Nonlinear systems are capable of displaying complex behavior even if this is the result of a small number of interacting  time scales. A widely studied case is when complex dynamics emerges out of a nonlinear system being forced by a simple harmonic function. In order to identify if a recorded time series is the result of a nonlinear system responding to a simpler forcing, we develop a discrete nonlinear transformation for time series based on synchronization techniques. This allows a parameter estimation procedure which simultaneously searches for a good fit of the recorded data, and small complexity of a fluctuating driving parameter. We illustrate this procedure using data from respiratory patterns during birdsong production.
\end{abstract}

\maketitle
\begin{quotation}
Nonlinear systems have the remarkable property that complex behavior can emerge out of seemingly simple rules. This invites the possibility that much of the complexity in a recorded time series can be explained by simple nonlinear mechanisms. Here we are interested in the case that the dynamics corresponds to a simple nonlinear system forced by an external signal. In order to test for this possibility, we define a procedure which aims at approximating the dataset with minimal complexity of the forcing signal. We test this approach in a model for the firing rate of populations of neurons with both model generated data and physiological recordings of the air sac pressure of a singing bird. We find model parameters for which the solutions of the forced system are a good approximation to the data and at the same time, the forcing signal exhibits low dynamical complexity. Thus, we show that there are parameters in the model for which representations of the dataset are encoded into simpler signals. 
\end{quotation}

\section{Introduction} 

Parameter estimation in smooth dynamical systems from experimental time series is central to many disciplines across the natural sciences. In many situations, the system under study can be modeled by a set of differential equations which governs the evolution of the state variables. Usually, such dynamical models will contain unknown parameters which one expects to estimate from the data. While much progress has been made in the case of linear dynamics, less is known about the more general case in which the dynamics involves nonlinear terms. Parameter estimation in nonlinear models is strongly case dependent partly because the model solutions can show a qualitative dependence on parameter values \cite{gucken}. Also, analytic solutions for a general smooth nonlinear system are typically unknown, posing serious difficulties when trying to associate estimated parameters with probabilities or likelihoods \cite{lutkepohl05}. 

It is widely believed that the temporal evolution of macroscopic phenomena obeys deterministic rules. Sometimes these rules are derived axiomatically as it is the case for classical and quantum mechanics, while in other areas of knowledge the proposed relationships attempt to capture phenomenological observations. In many situations, macroscopic phenomena is thought to emerge out of the collective action of many microscopic units, whose dynamical rules can be derived from \emph{ab-initio} principles. However, establishing correspondence between these two scales is one of the oldest and hardest challenges in physics. This difficulty becomes explicit when studying systems that operate far from equilibrium, as is the case for example, in biological systems. Independently of how these rules are derived, the problem of approximating data with solutions of differential equations seems to be an unavoidable step when comparing theories against real world data. This is especially difficult if the dynamical equations contain nonlinear terms of their state variables, even if the system is low dimensional. This problem has been addressed by many authors with particular emphasis in the case that the underlying dynamics is chaotic \cite{abarbanel1993, baker}. 

In many situations the causal relationships between the state variables of a system can be modeled by a set of differential equations or vector fields. Usually, the equations contain parameters which are assumed either to be stationary or to have a much slower temporal evolution than the state variables. Here, we consider such parameterized families of smooth vector fields. Let $F(x,p) : \mathbb{R}^{m \times q} \rightarrow M \subset \mathbb{R}^m$ be a smooth vector field with $x \in \mathbb{R}^m $ and $p \in \mathbb{R}^q$. We then have a continuously p-indexed set of solutions $\phi_t(x,p): \mathbb{R}_{ > 0} \times \mathbb{R}^{m \times q} \rightarrow \mathbb{R}^m$ for the associated initial value problem

\begin{eqnarray} 
\dot{x} &=& F(x,p)           \nonumber \\
x(0) &=& x_0
\label{ode}
\end{eqnarray}

Given a time series $o(t): [0,T] \rightarrow \mathbb{R}^d$ we can compare solutions against data using a metric $D: \mathbb{R}^m \times \mathbb{R}^d \rightarrow \mathbb{R}$. Let $(F,D)$ be a model for $o(t)$, we then investigate the following questions: \\

\begin{enumerate}
\item Is the proposed model able to reproduce/explain the data to desired accuracy? 
\item Which are the parameters that yield the best approximation? 
\end{enumerate}

Admittedly, there are more questions to be addressed such as robustness and likelihood of the approximation, but these can be regarded as ulterior inquiries once the answers to the first two points are found. One could attempt an answer by computing the error between the data and the model solutions as a function of the initial conditions and the parameters. The cost function can be defined as

\begin{equation}
\label{eq:contcostfunction}
C(x_0, p) = \frac{1}{T} \int_0^{T} D(\phi_t(x_0,p), o(t)) dt. 
\end{equation}

The functional form of $\phi_t(x_0,p)$ is in general unknown. Assuming that $\{o_i\} \approx o(t)$ is a discrete sampling of the data with $N$ elements, we consider a discretized cost function

\begin{equation}
\label{eq:disccostfunction}
C(x_0, p) = \frac{1}{N} \sum_{i=0}^N D(x_i, o_i).
\end{equation}

Here $x_i = \phi_{t_i}(x_0,p)$ are the values of the solutions in a discrete time domain: this function can be computed numerically to desired accuracy. Minimization of this function yields sets of initial conditions and parameters for which the solutions of the model better resemble the experimental time series, as measured by $D$. However, it is known that this  approach has severe limitations: there are no known global optimization methods for arbitrary nonlinear functions. Good solutions can be obtained efficiently by heuristic solvers such as genetic algorithms, simulated annealing, or the amoeba algorithm \cite{holland,kirkpatrick,neldermead}: however, finding the \emph{best} solution seems to be an intractable problem for many nonlinear systems. 

A major contributing factor to this intractability is that the landscape of $C$ can be extremely complicated. For the sake of illustration, let us assume that $\{o_i\}$ is the output of a computer running a simulation of a known low dimensional chaotic system $F_c$ and also that we have access to the program being run. Since the dynamical rules are known along with the initial conditions $x^*_0$ and the parameters $p^*$ that generate the data, we can check that the global minimum of $C(x^*_0,p^*)=0$, thus enabling us to conclude that we have the right model for the data, as expected. One validation criteria which we could posit would read as follow: there are points $(x^*_0,p^*)$ for which the model solutions resemble the data optimally, since lower errors cannot be attained. Although this criteria seems sufficiently general and does not require knowledge of $F_c$ solutions, in a more realistic setting, the value of the global minimum is unknown and all that can be addressed numerically is local information of the landscape. This information is completely useless even in this numerical \emph{gedankenexperiment}. 

Consider now what happens to the landscape of $C$ as more data is introduced. Imagine we are trying to estimate the initial conditions which were used to start the simulation: due to chaos, nearby solutions will diverge exponentially yielding high values in C. Eventually as $N$ is increased, the values of $C$ become settled at some average value resembling the size of the attractor in phase space. This happens for every point in the domain of $C$ that is not the global minimum. Therefore, the landscape near the global minimum becomes extremely sharp as more data is introduced posing a severe limitation to address our agenda.

This point is illustrated in Figure 1 for the Lorenz system \cite{lorenz}. We generated data by numerical integration of the Lorenz equations for values of the parameters such that there is a chaotic attractor (see figure caption). For this we picked initial conditions and parameters $(x^*_0,p^*)$ and simulated the system for $N$ steps. The resulting solutions are used as data $\{o_i\}$ for the cost function of the system (\ref{eq:disccostfunction}): we compute mismatch of model output and data with the least squares metric $D=(x_i-o_i)^2$, and assume that each $x_i$ corresponds to a time step in an Runge-Kutta O(4) integration routine with $dt=0.01$.  Figure 1 shows how the landscape of $C$ changes as $N$ is increased. When the number of data points is low, the global minimum is easily identified. As more data is introduced, the landscape near the global minimum becomes extremely sharp until it can no longer be resolved in the plots. The choice to evaluate $C$ as a function of initial conditions was made for simplicity but the same properties should hold for parameter space. Because the attractor persists for small variations of the parameters, a similar argument can be made with the same outcome. For values of the parameters close to $p^*$, the shape of the attractor changes but the mixing properties of the vector field remain invariant: if initial conditions are kept fixed and a neighborhood of $p^*$ is explored, solutions will land on attractors that are slightly different and this will have the same effect as if the same attractor is being approached from slightly different initial conditions. Overall, this example suggests that if we are to test the hypothesis that data corresponds to a chaotic system, a departure from the cost function is in order.  

%\begin{widetext} 
\begin{figure*}[htb!]
\label{fig:lorenz}
\includegraphics[width=160mm]{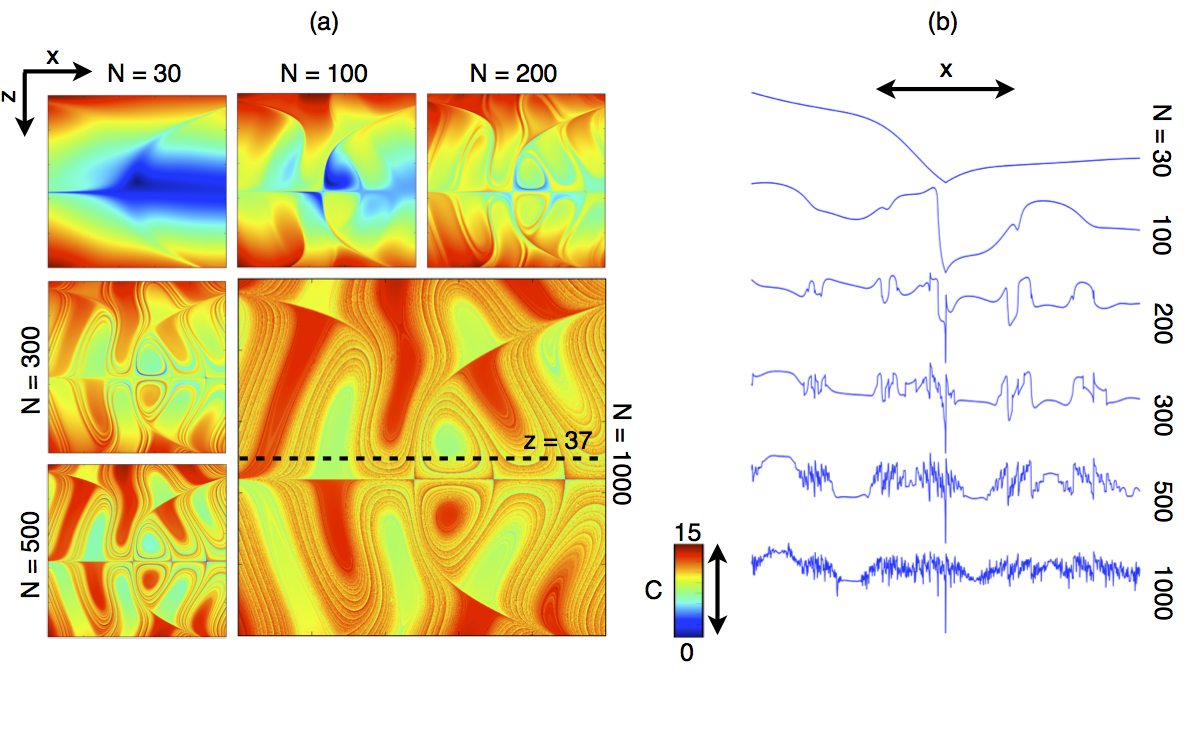}
\caption{ Evaluation of the cost function for the Lorenz system: $(\dot{x},\dot{y},\dot{z})= (\sigma ( y-x),x ( \rho - z) - y , xy - \beta z)$ with $p^* = (\rho = 28 , \sigma = 10 , \beta = \frac{8}{3})$. A test solution is generated using a Runge-Kutta O(4) routine ($dt=0.01$) with initial conditions on the attractor $x^*_0 = (-10,0,37)$ and used as data for the cost function in a sample domain consisting of a regular 1000x1000 grid.  The plotted domain is $(x_0,z_0) = [ (-90,70), (-43, 117) ]$ and it was chosen so that the test solution corresponds to the center of the plots. (a) Each panel corresponds to the landscape of the cost function for the Lorenz system as the number of data points $N$ is increased. The cost function value is color coded (color scale shown in plot). In all panels, the initial conditions of the test solution are in the center of the plot and correspond to the global minimum of this function C(-10,73)=0. (b) Unidimensional plot of the cost function for $(x_0, z_0=37)$ indicated in (a) by the dashed line. Note that for low N, the minima of this function is easy to find: however, as N is increased, the landscape surrounding the global minimum becomes extremely sharp and therefore, information about the initial conditions of the test solution can no longer be retrieved from the 2D plots. }
\end{figure*}
%\end{widetext}

The keen observer has probably realized that if  the number of data points $N$ is low, the minima looks round and is easy to find, suggesting that tracking the parameters over short segments of time could prove useful. This is the idea behind coupling the model to the data and it has been explored by many authors: in particular, it has been show that under certain conditions, when the model is coupled to the data the cost function is smoothed or `regularized', and parameters can be estimated \cite{konnur, parlitz, parlitzprl}. Unfortunately, the regularized cost function can be extremely flat and therefore, finding its minima becomes hard to do numerically. Recently, Abarbanel et al. proposed a method to estimate parameters in chaotic models called DPE \cite{abarbanel2008,abarbanel2009} which addresses this issue. They regularize the cost function by coupling the model with the data while penalizing coupling by adding a term proportional to the coupling strength in a new balanced cost function: this leads to a function that can be minimized and yields the right answer. We follow a closely related approach, but aim for a different question: can data be approximated by smooth unfoldings of low dimensional systems? 

Synchronization-based parameter estimation techniques rely on coupling the data to the model equations and are at the heart of many estimation procedures \cite{sorrentino2009}. In this work we introduce a synchronization-based transformation for time series which aims at exploring the possibility that data can be approximated by a driven nonlinear system. For this we explore a family of synchronization schemes by allowing model parameters to change in time so that the resulting solutions would better resemble a given dataset. In this way  a map is induced between transient solutions of the model and the parametric fluctuations which are required to generate them. We assume that in general, better approximations come at the cost of introducing more complex fluctuations. However, it could be the case that  there are parameters for which the fluctuations take a particularly simple form at the same time that they yield solutions which are good approximations to the dataset. 

In what follows, the transformation is motivated by considering an idealized continuous case while rigorous definitions are postponed to the next section. In order to allow for better approximations we consider an extension of the cost function which includes arbitrary parametric fluctuations in $F$. Let $\gamma(t): [0, T] \subset \mathbb{R} \rightarrow W \subset \mathbb{R}^{q}$ be a path in parameter space with $W = p \pm \epsilon$. For each $(x_0,p)$ we can define the functional $\mathbb{E}_{x_0,p}: {\gamma(t) \times o(t)}  \rightarrow \mathbb{R}_{\geq 0}$ as

\begin{equation}
\label{eq:functionalcost}
\mathbb{E}_{x_0, p}[\gamma(t), o(t)] = \frac{1}{T} \int_0^{T} D(\phi_t(x_0,p + \gamma(t)), o(t)) dt.
\end{equation}

Let $\textbf{e}(x_0,p) \in \mathbb{R}$ be the minimum value of $\mathbb{E}$ and $\gamma^*(t)$ the `location' of the minimum, $E_{x_0,p}[ \gamma(t), o(t) ]=\textbf{e}(x_0,p)$. We define the nonlinear transformation of our dataset as 

\begin{eqnarray} 
\Omega_{x_0,p} [ o(t) ] = \gamma^*(t). 
\end{eqnarray} 

Ideally, this transformation takes the experimental time series $o(t)$ as input and returns the \emph{fluctuations} $\gamma(t)$ in the space of parameters that are necessary for system (\ref{ode}) solutions to optimally match the dataset as measured by $D$. The goal here is to define a map between a particular transient solution of the model and the parametric fluctuations which generate it. Since parameter fluctuations greatly augment the degrees of freedom of $F$ it is expected in general that $\textbf{e}(x_0,p)  \ll C(x_0,p)$. For a sufficiently complex fluctuation $\gamma(t)$ we will be able to approximate $o(t)$ very well regardless of the structure imposed by our assumptions. We note however, that for different values of the parameters of $F(x,p)$, there might be other simpler fluctuations which might approximate $o(t)$ similarly well. Our objective is to select the parameters for which the optimally matching fluctuations take the simplest form while the approximation also meets a goodness criteria. Let $K[\gamma(t)]: \gamma(t) \rightarrow \mathbb{R}$ be a measure of the complexity of the required fluctuations such that $K[\gamma(t)]=0$ when $\gamma(t)=0$. Our hypothesis is that for most parameter values there is a tradeoff between $e(x_0 ,p)$ and $K[\gamma(t)]$ in such a way that obtaining good solutions comes at the cost of increasing the complexity of the corresponding fluctuations. Therefore, we seek values of the parameters for which our hypothesis does not hold: namely, better approximations are obtained by simpler parametric fluctuations. If the fluctuations that are required to approximate a given dataset are as complex as the dataset itself, we would argue that the model is acting as a complicated mapping. However, if the approximations are generated by simple time dependent parametric fluctuations, any additional complexity that the data might have which is captured by the model solutions, would be due to \emph{specific} nonlinearities in $F$.  If it is true that there is a negative correlation between the error of the approximations and the complexity of the corresponding fluctuations, departures from this general tendency could be quantified and exploited to detect parameter values of interest. 

In the next section we propose a discrete nonlinear transformation for time series inspired by the continuous case. Later in this article this definition is applied to construct criteria for model selection in a hypothetical scenario: rate coding in driven neural populations. It is shown numerically for a test case that $log(e(x_0,p)) \approx m K[\gamma(t)] + b$, for a particular choice of $K$.  This in turn is exploited to estimate the parameters and the driving signal used to generate the data. The same construction is finally applied to experimental data.  

This work is organized as follows: section 2 contains the definitions and describes a numerical implementation of the transformation. In section 3 we analyze a case study: a simple model for the rate of fire in neural tissue forced by an external signal. We test the usefulness of the transformation with numerical solutions generated by the model. It is shown numerically that we can identify the parameters used to generate the data without making assumptions on the functional form of the forcing signal. Finally, the problem of motor patterns in birdsong is discussed in section 4. We show that there are regions in parameter space for which the synchronization manifold of the model and the respiratory gestures used during song production exhibits subharmonic behavior.  

\section{Definitions and numerical implementation} 
In this section we discuss a numerical procedure inspired in the previous discussion. We assume that data is sampled at discrete time intervals and that $F(x,p)$ satisfies the hypothesis of a Runge-Kutta O(4) method \cite{numericalrecipes}.  Let $\{o_i\}$ be a time series with $N$ samples and $\{x_i(x_0,p)\}$ be a numerical solution of

\begin{eqnarray}
\label{eq:coupledF}
\dot{x} &=& F(x,p+\gamma(t)) \\ \nonumber
x(t=0) &=& x_0 \\  \nonumber
\gamma(t=0) &=& \gamma_0.
\end{eqnarray}

Our focus is to find $\gamma(t)$ such that $\{x_i\}$ matches $\{o_i\}$ as measured by $D$. For simplicity, we assume that each observation corresponds to a time step of the numerical approximation $\{x_i\}$. In order to define the transformation $\Omega$ we specify a domain for the fluctuations. Let $\epsilon$ be the maximum allowed departure from $p$ so that at each integration step $\gamma$ can take any value in the range $W = (p-\epsilon, p+\epsilon)$. For each $(x_0,p)$, the cost function of the system (\ref{eq:disccostfunction}) can be augmented so that it includes parameter fluctuations at each integration step,

\begin{equation}
\label{eq:highdimcost}
E_{x_0, p}(\{ \gamma_i\}) = \frac{1}{N} \sum_{i=0}^{N} D(\phi_{t_i}(x_0,p+\gamma_i), o_i).
\end{equation}

In this way the functional optimization (\ref{eq:functionalcost}) is approximated by a high dimensional nonlinear optimization problem without constraints: while in general this optimization cannot be performed for large $N$, there are efficient approaches for particular choices of $F$ and $D$. 

We define the transformation $\Omega_{x_0,p}(\{o_i\})$ as

\begin{equation}
\label{eq:omegadefinition}
\Omega_{x_0,p}(\{o_i\}) = \argmin_{\{\gamma_i\}}(E_{x_0, p}(\{ \gamma_i\})),
\end{equation}

and introduce the notation for the corresponding model output

\begin{equation}
\Omega^{*}_{x_0,p} (\{\gamma_i\}) =\vcentcolon \{ \phi_{t_i}(x_0,p+\gamma_i) \} = \{x_i\},
\end{equation}

where the sequence $\{\phi_{t_i}(x_0,p+\gamma_i)\}$ is a numerical approximation of the non autonomous system (\ref{eq:coupledF}) given by a Runge-Kutta (O4) method \cite{numericalrecipes}. Here $x_0$ and $p$ are fixed and our aim is to solve for $\{\gamma_0 . . . \gamma_N \}=\text{argmin}(E)$. The minima of this function corresponds to the fluctuations $\{\gamma_0 . . . \gamma_N\}$ that yield the observable time series $\{x_0 . . .x_{N} \}$ which better approximates the data.
 In order to tackle the optimization problem the fluctuations are restricted to take a discrete set of possible values. Let $R$ be the number of equally spaced values that $\gamma_j$ can take at each integration step. The number of possible solutions is then $\propto R^N$, thus, attempts to optimize this function are bound to fail for large $N$. 

The strategy we then follow is to calculate $\{ \Omega_i \}$ in small running windows of size $d<<N$. Since in general we cannot minimize nonlinear functions, we perform this task by brute force search on a discrete set of $R$ equally spaced values in $W$. In any given window, this yields the model solution $\{x_0, . . . , x_d\}$ and the fluctuations $\{\gamma_0,  . . . , \gamma_d\}$. We define a new $\Omega$ for the subsequent window using the first step of the approximation $x_1$ obtained at step $i$.  Finally we define the transformation $\Omega^{d,R}$, 

\begin{eqnarray}
\gamma_{i+1} &=& \{ \Omega_{x_i,p}[ {o_i . . . . o_{i+d}}] \}_{1} \\ \nonumber
x_{i+1} &=& \{ \Omega^{*}_{x_i,p} [\Omega_{x_i,p}[ {o_i . . . . o_{i+d}}]] \}_{1}.
\end{eqnarray}

By iterating this map two time series are obtained: the solution $\{x_i\}$ of system (\ref{eq:coupledF}) that aims to approximate the data and the fluctuations $\{\gamma_i\}$ which are required to generate it. Thus, for each point in $(x_{0} , p)$ we have defined a three parameter transformation $(\epsilon,R,d)$ of the dataset. This is a special case of a more general definition described in the appendix. 

In the next section this transformation is applied to a case study. We devise a scenario in which complex output is attained by simple parametric oscillations of a low dimensional dynamical system. It is found that this procedure yields a good approximation for $\Omega$. This is supported by numerical results in the next section but it is not expected to be the general case. However, we do expect that if we have the right model, it will synchronize with the data and good approximations will be obtained. In general we expect that by allowing parametric fluctuations, better approximations can be achieved, even if the model is wrong. In that case, we also expect that the required fluctuations are complex. The purpose of this procedure is to systematically investigate the possibility that there are parameters in the model for which this transformation takes a simple form while still approximating the data. 

\section{Case study: Rate coding in populations of neurons.} 

\begin{figure*}[hbt!]
\label{fig:dataset}
\includegraphics[width=160mm]{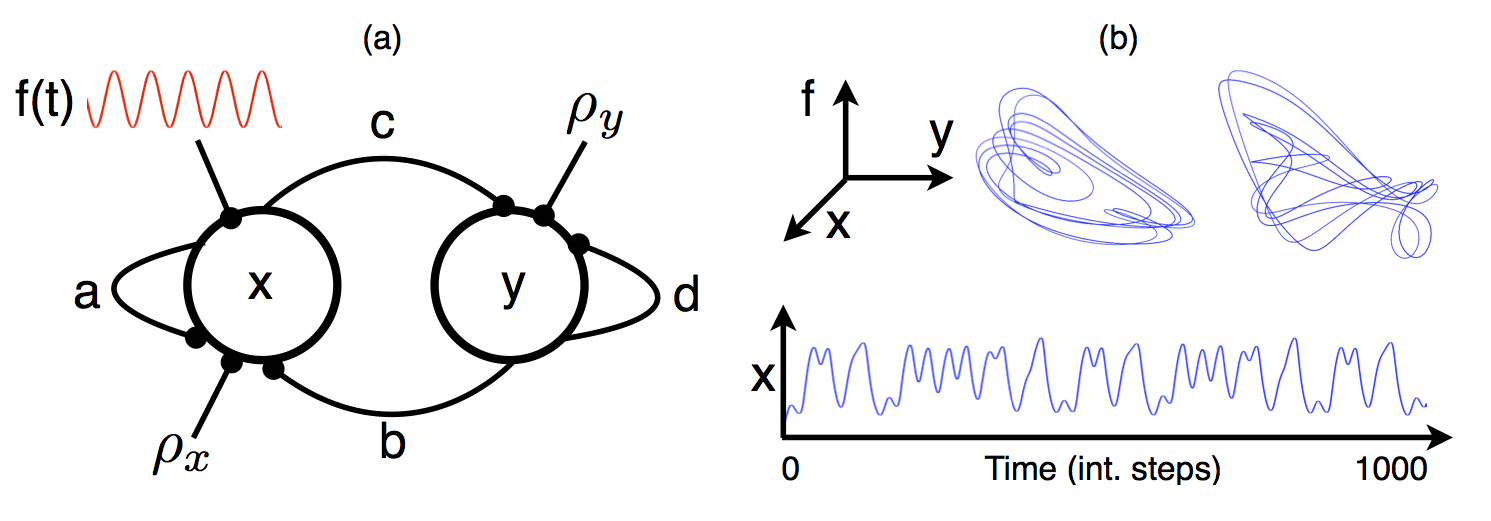}
\caption{Data was generated by numerical integration ($dt=0.1$) of eqns. (\ref{eq:forcedWC}). Small changes in the harmonic forcing parameters $(A,\omega)$ yield wildly varying responses ranging from very simple quasi harmonic oscillations to apparent non periodic behavior.(a) Schematic representation of system (\ref{eq:forcedWC}). (b) Full solution in phase space and projection onto the x coordinate for $(A = 1.2, \omega=2.1)$.}
\end{figure*}

In this section we study a phenomenological model for neural tissue under different perturbation regimes. The assumption is that the firing rate of a population of neurons decays exponentially in time in the absence of stimuli and that the population response to input becomes saturated after a threshold value. Here we study a simple architecture consisting of two interacting populations of inhibitory and excitatory neurons.  This is also known as the additive model and it was first proposed as a model for populations of localized neurons by Wilson and Cowan \cite{wilsoncowan72},

\begin{eqnarray}
\label{eq:wc}
\dot{x} &=&  ( -x + S(a x - b y + \rho_x) ) \tau\\ \nonumber
\dot{y} &=&( -y + S(c x - d y + \rho_y) )  \tau.
\end{eqnarray}

An excitatory population is coupled to an inhibitory one through the connectivities $(a,b,c,d)$, both populations receive a constant input given by $(\rho_x,\rho_y)$ and $\tau$ is the timescale of the system. The assumption of saturation is implemented by the sigmoid function $S(x)= \frac{1}{1 + e^{ -x}}$. This system presents a rich bifurcation diagram in the plane $(\rho_x, \rho_y)$ for open sets of values of the connectivities \cite{weakly}. Since the system is two dimensional, we can't expect its solutions to be more complicated than a limit cycle. However, if we also assume that the system is being driven by an external signal a wild diversity of behaviors are unleashed.

In order to test the transformation introduced in the last section, a dataset $o(t)$ was generated by driving the model with simple fluctuations in parameter $\rho_x \rightarrow \rho_x + \gamma(t)$. In this scenario, $\gamma(t)$ is interpreted as an external signal that is being injected into the excitatory population and processed by the network. We are interested in testing our hypothesis for the case in which complex output is attained by simple parametric fluctuations: in this spirit, we explore a family of fluctuations parametrized by $\gamma(t)=A \cos(\omega t)$, 

\begin{eqnarray}
\label{eq:forcedWC}
\dot{x} &=& ( -x + S(a x - b y + \rho_x + A \cos(\omega t)) )  \tau\\ \nonumber
\dot{y} &=& ( -y + S(c x - d y + \rho_y) ) \tau.
%\dot{\phi}&=& \tau ( \omega)
\end{eqnarray}

We choose parameters so that the system undergoes bifurcations as $\gamma(t)$ evolves: $(x_0,y_0) = (0,0)$ and $(a,b,c,d,\rho_x,\rho_y,\tau) = (10,10,10,-2,-3,-8,1)$. System (\ref{eq:forcedWC}) is then numerically integrated using a Runge-Kutta O(4) routine with temporal resolution $dt=0.1$. For this choice of parameters the global attractor of the \emph{autonomous} system ($A=0$) is an attractive limit cycle:  a stable nonlinear oscillatory solution. It is know that these solutions can synchronize with an external periodic forcing \cite{synchronization}. Due to the nonlinear nature of these solutions, this entrainment occurs in very different ways: for some regions of $(A,\omega)$ space the period of the solution matches the forcing period (1:1), but there are regions for which this ratio can take in principle any (p:q) value. These regions are generically known as Arnold tongues due to their V-shape in $(A,w)$ space and we refer to them collectively as the Arnold tongues diagram of the system \cite{arnold}. The morphology of the solutions greatly differs from one tongue to another so this offers an appealing strategy for coding different morphologies into simple control signals. While the quantitative features of this organizing map are model dependent, the qualitative geometrical mechanisms by which these solutions are generated can be characterized by topological methods\cite{classification, alicestretchland}. A schematic representation of the model is shown in Figure 2 (a). 
Figure 2 (b) shows the dataset to be analyzed in this section. This solution was found by computing numerically the Arnold tongues diagram of the system in $(A,\omega)$ space and searching for responses with a period larger than 10 times the period of the forcing. 

%\begin{widetext} 

%\end{widetext} 

Next we ask if we can recover the test parameters by assuming the unperturbed system (\ref{eq:wc}) and generic input,

\begin{eqnarray}
\label{eq:wcgenforcing}
\dot{x} &=&  ( -x + S(a x - b y + \gamma(t) ) ) \tau \\ \nonumber
\dot{y} &=&  ( -y + S(c x - d y + \rho_y) ) \tau \\ \nonumber
\dot{\gamma} &=&  (\rho_x - \gamma) \tau_{\gamma}.
\end{eqnarray}

The role of the third equation is to force the input to population $x$ to have bounded derivative. This hypothesis is explicitly included in the model equations via the auxiliary variable $\gamma(t)$ and parameter $\tau_{\gamma}$. Note that if $\gamma(t=0)=\rho_x$, solutions of system (\ref{eq:wcgenforcing}) are identical to those of system(\ref{eq:wc}). We explore coupling of this augmented model with data by computing $\Omega[o(t)]$ when parameter $\rho_x$ is allowed to fluctuate. The effect of the additional equation can be roughly described as follows: for high values of $\tau_{\gamma}$ we'll have that $\rho_x(t) \approx \gamma(t)$, while for lower values of $\tau_{\gamma}$ we'll obtain smoother oscillations in $\gamma(t)$. Here we note that while the resulting parametric fluctuations $\{\rho_{x_i}\}$ represent a candidate solution to eqn. (\ref{eq:omegadefinition}) when assuming system (\ref{eq:wcgenforcing}), the corresponding solution $\{ \gamma_i \}$ also represents an approximation to eqn. (\ref{eq:omegadefinition}) when considering the unperturbed system (\ref{eq:wc}). Since both alternatives yield the same error value,  we can choose to consider data as being approximated by system (\ref{eq:wc}) when forced by $\{ \gamma_i \}$ or as being approximated by system (\ref{eq:wcgenforcing}) when forced by $\{ \rho_{x_i} \}$. This choice will become important when quantifying the complexity of the resulting transformation due to the choice of $K$ and also when considering experimental data.

%However, if fluctuations in parameter $\rho_x$ are allowedThe model and the data 

Let us summarize the scheme adopted for the rest of the article: parameter fluctuations are allowed in system (\ref{eq:wcgenforcing}) by coupling model and data through $\rho_x$. The transformation was done as depicted in the previous section: at each time step, the fluctuations take any of $R=200$ equally spaced values in $p=(a,b,c,d,\rho_x,\rho_y,\tau,\tau_{\gamma}) \pm \epsilon = (0,0,0,0,5,0,0,0)$. Window size for local fits is $d=2$. This process yields an approximating time series $\{x_i\}$ along with the fluctuations $\Omega_{x_0,p}[\{o_i\}]=\{\rho_i\}$ that generate it. As discussed before, we can consider that $\Omega_{x_0,p}[\{o_i\}]=\{\gamma_i\}$ provided we state that system (\ref{eq:wc}) is being assumed. Finally, the metric by which model approximations and data are compared is $D= (x_i-o_i)^2$ where each $x_i$ corresponds to a time step in the integration routine. 

Our aim is to characterize the parametric dependence of  $\Omega_{(x_0,p)}[ \{o_i \}]$. For this we perform the transformation $\Omega_{x_o,p}[\{o_i \}]$ in a sample domain consisting of $n=1001$ equally spaced points in parameter space along coordinate $\tau \subset (0,2)$. The rest of the parameters were kept fixed at $p=(10,10,10,-2, -3 \pm 5,-8,\tau,10)$ as well as the initial conditions $(x_0, y_0, \gamma_0)=(0,0,\rho_{x_0}=-3)$. The point $\tau = 1$ was purposely excluded from the sampled domain as a test for robustness of the results. Figure 3 panel (a) shows the approximations to dataset $\{ o_i\}$ along with the transformation (b) $\Omega[ \{ o_i\}  ]$ we obtained for values in a neighborhood of $\tau=1$. This figure summarizes the main result of this article: the quality of the approximations is very similar for most values of $\tau$, however, for $\tau=1.05$ the fluctuations take a particularly simple form, which resembles very much the harmonic forcing used to generate the test data. Thus, we have produced a parameterized family of transient solutions which intersect the test solution. We find that within this parametrization there are bold variations on the dynamical complexity of the resulting transformation. This observation was found to be robust for many choices in the transformation parameters $(\epsilon, d,R)$. 

%\begin{widetext} 
\begin{figure*}[hbt!]
\label{fig:transforms}
\includegraphics[width=140mm]{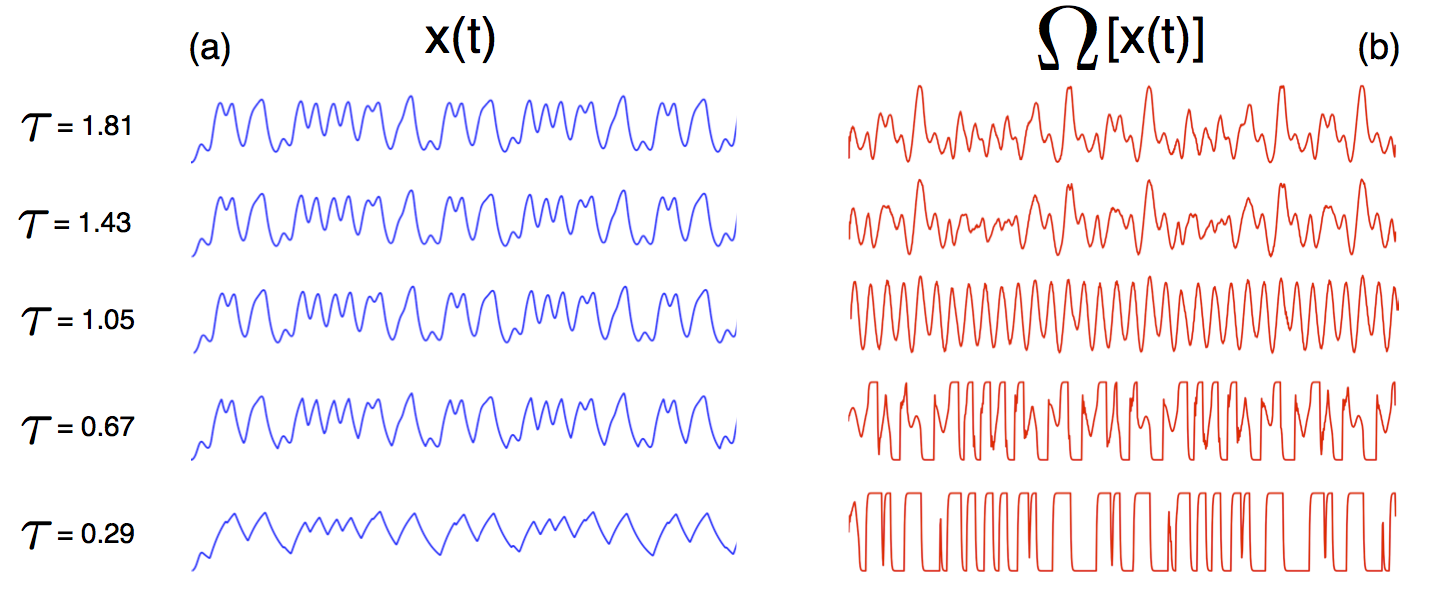}
\caption{Dependence of $\Omega_{x_0, p}[\{o_i\}]$ on model parameters. Test data is approximated by model (\ref{eq:wcgenforcing}) at the cost of introducing fluctuations in parameter $\rho_x$. For most values of $\tau$ good approximations are obtained, $x_i \approx o_i$. However, the complexity of the required fluctuations shows a marked dependence on the model parameters. Note that for $\tau=1.05$ the transformation takes a very simple form, thereforeparsimony suggests that values close to $\tau \approx 1$ should be chosen. (a) Best approximations to the dataset obtained values of $\tau$ indicated in labels. (b) Fluctuations required to generate the approximations. For each solution $x(t)$ in the right, we show $\Omega_{x_0,p}[x(t)]$ in the left: correspondence follows from definition (\ref{eq:omegadefinition}).}
\end{figure*}
%\end{widetext} 

In order to quantify the complexity of the fluctuations we calculated their permutation entropy $K$  \cite{perment}. This quantity measures the information of the order type distribution of $N_K$ consecutive values of a time series. Since the number of possible order types is $N_K!$, the order of permutation entropy has to be low enough as to assure that the order type distribution is reasonably sampled. In this case, since the number of data points is $N=1000$ the permutation entropy was calculated to order $N_K=5$. The quality of the approximation can be captured by the error as defined before (\ref{eq:highdimcost}),
 
\begin{equation}
  e = E_{x_0,p}(\{\gamma_i\}).
\end{equation}

Figure 4 (b) shows the values of error $e$ and entropy $K$ of the fluctuations as a function of $\tau$. While the error decreases monotonically, the entropy curve exhibits a dent. This means that, for the tested domain, as error decreases entropy increases except for the dented region, where this simple relationship seems to break. To make this statement  more precise, Figure 4 (a) shows the error-entropy distribution obtained by evaluation of the transformation in the sample domain. We quantify the aforementioned tradeoff by fitting a line ($\log(e) = m K + b$) to the cloud and coloring the points according to their normal distance to the fit. In this way departures from the negative correlation are addressed quantitatively. For each point in $(\log(e), K)$ the normal intersection with the fitted line occurs in $K_i = \frac{K +m(\log(e)-b)}{(m+m^2)}$, $\log(e)_i = m K_i +b$. The distance to the fit then reads 

\begin{eqnarray} 
\label{eq:H1}
d(x_0,p) &=& \sqrt{(\log(e)-\log(e)_i)^2 + (K-K_i)^2} \\ \nonumber
              &\text{  }& \times \text{    sign}(\log(e)-m K - b),
\end{eqnarray} 

where a $sign$ function was introduced to differentiate between favorable departures (below the line) from unfavorable ones. Departures above the fit correspond to solutions that should fail the identification criteria. This is either because the approximation is not good enough or more importantly, because the fluctuations required to generate them exhibit high dynamical complexity. By seeking minima of $d$ it can be assured that within the family of approximations induced by $\Omega^{2,200}$, the most parsimonious one in the sense of $K$, is chosen. This enables a \emph{lex parsimoniae} criteria within the restricted family along with a proof of existence of possibly interesting mechanisms. 

Figure 4 (a) shows the plot of $d$ over the test domain (green curve). This quantity compares the quality of the approximations with the complexity of the required fluctuations and it is statistically defined through $m$ and $b$. In this case, the parameters used to generate the data can be estimated by searching for minima of $d$. These are the parameters for which a favorable balance is obtained between goodness of the approximation and the complexity of the extra assumptions contained in $\gamma(t)$. This allows us to define a parsimony criteria to estimate the parameters of system (\ref{eq:wc}) which in this case leads to the correct answer $\tau \approx 1$. This suggests that this mapping can be useful to create objective functions to train dynamical models toward simpler representations of a given dataset.

In order to make the result somewhat more general we could argue as follows: we can assume that in general there is a relationship between the argument and the minimum value of function (\ref{eq:highdimcost}) of the form $log(e^*) \approx m K[\Omega] + b$. Then we can state that any sequence $\{ \gamma_i \}$ which violates the scaling is of interest and leave the matter of how these sequences are generated free of assumptions. Our argument then proceeds as follows: at any point in the sample domain $(x_0,p)$, we have attempted an approximation to the problem of computing $\Omega$ by trying different parameters in $\Omega^{d,R}$. By further assuming there is $\delta$ such that if $|e^* - e|<\delta$ then $K[\Omega] \approx K[ \Omega^{d,R} ]$, we may regard $\Omega^{d,R}$ as a good approximation to $\Omega$. This extends the parsimony rule by comparing all possible paths within numerical resolution. Plausibility of this last assumption is supported \emph{a posteriori} by the numerical results in this case study. However, in the experimental case, this assumption is more daring since location of the global minimum of $E[\{\gamma_i\}]$, or alternatively $\Omega$, is always unknown.

While this function can help choose parameters within a sample domain, its evaluation requires knowledge of the error-entropy distribution: this makes it hard to implement an optimization routine over a search space. In a typical experimental setting, we could be interested in every solution which satisfies some goodness criteria. Amongst the many good approximations, interest is shifted towards those which require the simplest fluctuations. This can be implemented in a function: 

\begin{equation}
\label{eq:H2}
H(x_0,p) = (e + K_b)\Theta(e- \sigma) + K \Theta(\sigma-e),
\end{equation}

where $K_b$ is a bound for the permutation entropy $K \leq log_2(N_K!) \approx K_b=6.9$ (for $N_K=5$), $\Theta(x)$ is the Heaviside function and $\sigma$ is a threshold which represents a passing criteria for the approximations. Both arguments of the Heaviside functions cannot be true simultaneously. If the error is above the threshold, the function returns the error plus the entropy bound, otherwise, it returns the entropy value. Once the approximations drop below the threshold, the only way to obtain lower values in  $H$ is to find solutions that satisfy the threshold and require simpler fluctuations. This function corresponds to the red plot in Figure 4. In this case, the threshold is the mean value of the error over the sample domain $\sigma=<e>$.

\begin{figure*}[hbt!]
\label{fig:numericresults}
\includegraphics[width=160mm]{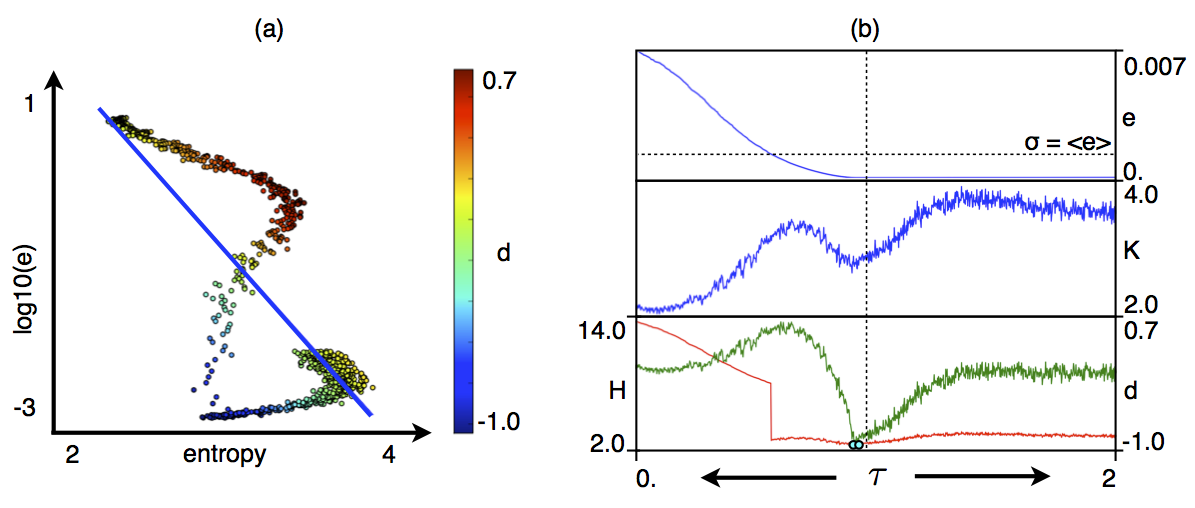}
\caption{Transformation $\Omega_{x_0,p}^{2,200}[\{o_i\}]$ was computed for each point in the sample domain. This yields the model approximation $\{x_i\}$ and the required fluctuations $\{\gamma_i\}$. Goodness of the approximation is quantified by the error $e$. The complexity of the fluctuations $\{\gamma_i\}$ is quantified by estimating the permutation entropy $K_5$. (a) Distributions of $log_{10}(e)$ and entropy $K_5[\{\gamma_i\}]$ with linear fit ($m=-1.85 , b=1.5 $). (b) Top row: error as a function of $\tau$. Center row: entropy as a function of $\tau$. Bottom row: (green) distance to the linear correlation fitted in (a) as a function of $\tau$. (red) Function (\ref{eq:H2}). While the general tendency is that lower errors correspond to higher entropies, the entropy curve exhibits a dent which breaks the scaling. These are consecutive points in the sampled domain for which the balance between error and entropy is favorable and also indicate roughly the parameter values which were used to generate the data $\tau=1$ (indicated by the dashed line).}
\end{figure*}

One would expect that if there are solutions such that the approximation is good and at the same time the fluctuations are simple, these will correspond to local minima of both functions for many choices of the sampling space and threshold. While the first one provides a tradeoff criteria to justify model parameters, the second one can be used as an objective function to train dynamic neural networks. If there is a negative correlation between error and entropy, a set of solutions that are simple and approximate data well would look like horns departing from the cloud that can be `shaved' by choosing different values of $\sigma$ in $H(x_0,p)$. 

Finally, it should be noted that the reason that both the model parameters and the forcing signal can be `guessed' at the same time is because of the fact that the fluctuations used to generate data were simple. Consider the case in which data is generated with complex fluctuations (for example, Fig. 3 with $\tau =1.5$). It would not be true anymore that transformations corresponding to nearby values of $\tau$ will be more complex than the fluctuations used to generate the data. While the error of the approximations will still be low for many choices of $\tau$, the entropy curve will still exhibit a dent around $\tau \approx 1$ and it would be wrongly concluded that data was generated with a harmonic forcing with $\tau \approx 1$. Another situation in which our approach fails is that in which data is generated by a system driven by noise, since in this case the real fluctuations would not be simple. It could also happen that there are no dents in the entropy curve. In this case it could be argued that since there are no parameters for which the fluctuations take a simple form, the structure of the model is spurious and that it should be discarded. In the next section we apply this ideas to a well studied example in motor control from the field of birdsong. 

\section{Case study: Complex motor patterns in birdsong.}

During song production, canaries rely on a repertoire of motor gestures which are ultimately responsible of driving their vocal organ. These gestures roughly correspond to the muscular activity by which the tension of the vocal folds is controlled and the respiratory activity which governs the air flow through the syrinx \cite{simplemotorgestures}. Here, experimental data consists of recordings of the air sac pressure of a canary while singing and it is shown in Figure 5(b) top row (green curves). We show that there are parameters for system(\ref{eq:wc}) such that the data can be approximated by model solutions if a simple forcing is also assumed.

In this example we chose a segment that contains a transition between two apparently different gestures: data consists of two approximately periodic signals with similar frequencies and marked morphological differences. Note that if no external forcing is assumed there is little hope that the 2 dimensional system (\ref{eq:wc}) will approximate the data. One can speculate that data can be approximated by the forced system (\ref{eq:forcedWC}) if the amplitude and frequency of the forcing is allowed to change when the gesture changes. We can include the possibility that the system is being generically forced by assuming system (\ref{eq:wcgenforcing}). This assumption greatly increases the goodness of the resulting approximations yielding remarkably good fits.

The objective is to check if there are parameters for which the model approximates data with simple fluctuations. A solution is considered `good' if it satisfies that the error is less than a threshold $\sigma = 0.002$. Once this condition is achieved, we seek to minimize the complexity of the required fluctuations, thus `increasing' the parsimony of our assumptions. This is done by optimizing function (\ref{eq:H2}) on a search space over parameter and initial conditions space. The reason we choose to minimize eqn. (\ref{eq:H2}) instead of eqn. (\ref{eq:H1}) is that evaluation of eqn. (\ref{eq:H1}) requires knowledge of the error-entropy distribution. Once a solution is found it can be checked if it is also a local minima of eqn. (\ref{eq:H1}). 

The optimization is performed by a genetic algorithm over a search space as follows: $x_i = 0.5 \pm 0.5$, $\rho_i = 0 \pm 20$, $(a,b,c,d)= 0 \pm 20$, $\tau = 2 \pm 2$ and $\tau_{\gamma} = 5 \pm 5$. In order to train the algorithm, we used a `low resolution' transformation by setting $R=2$. This allowed us to test for many solutions in a computationally efficient way. At this point we should point out that the landscape of $H$ will change dramatically on the choice of the parameters for $\Omega$. However, the suspicion remains that broad features of the landscape of $H$ are invariant to this parameter after some value, and it turned out to be a successful strategy for training the algorithm. The final scheme we adopted to define $H$ was $(d=2, R=10)$. The maximum amplitude for the fluctuations was also adjusted to $\rho_x \pm \epsilon = 0.85$ . Optimization was performed on a commercially available desk computer: the algorithm was run for 1000 generations starting from 100 random seeds and yielded parameters $p = (a,b,c,d,\rho_y,\tau, \tau_{\gamma}) = (3.03,3.17,13.65,-8.94,-12.64 ,1.53,9.15)$ and initial conditions $(x_0,y_0,\gamma_0) = (0.25,0.47,0.06)$.

\begin{figure*}[htb!]
\label{fig:experesults}
\includegraphics[width=130mm]{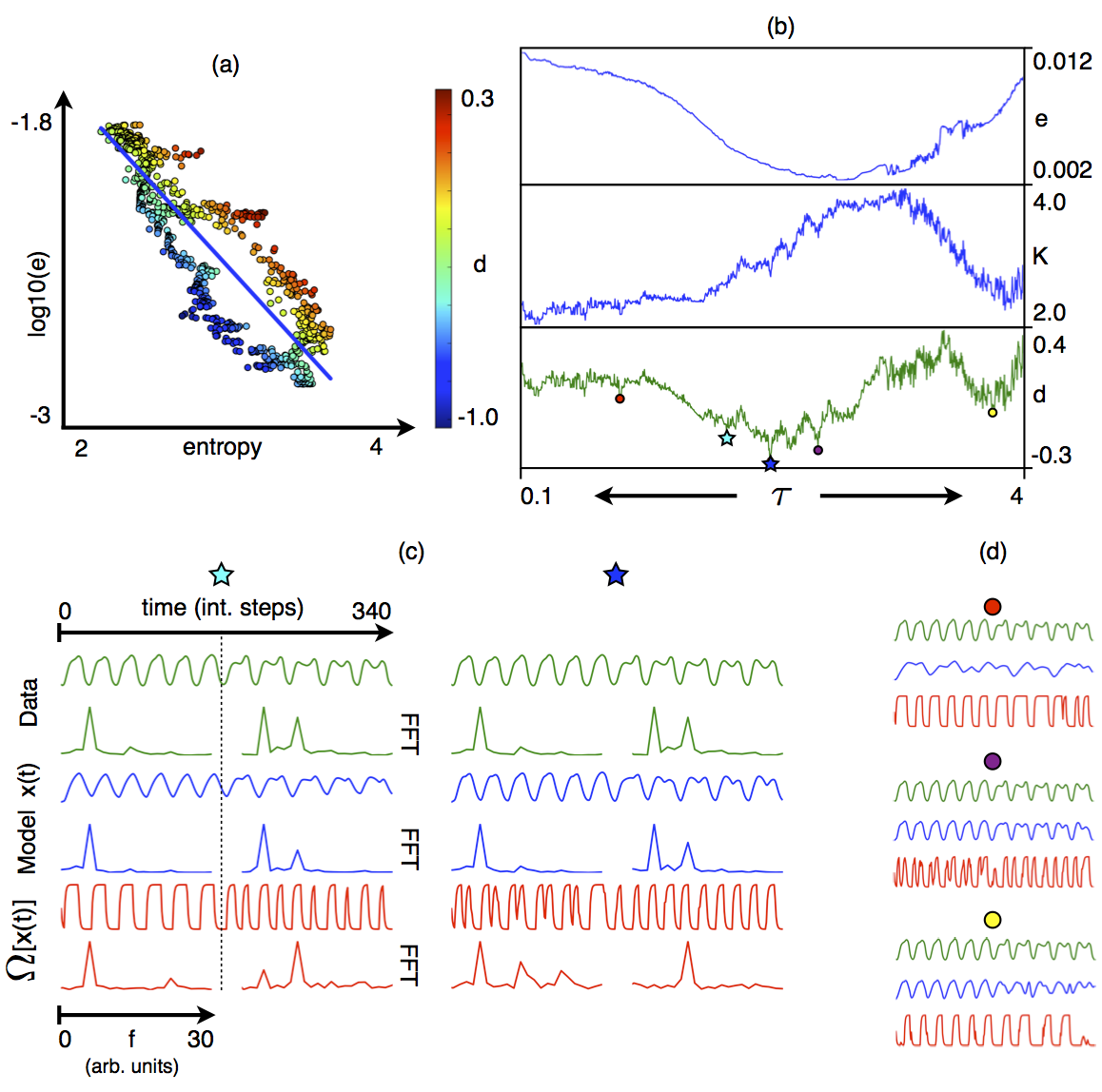}
\caption{Results on experimental data. Data corresponds to the air sac pressure of a canary while singing. An integration step in the model corresponds to $dt = \frac{1}{440} secs$ and the total duration of the recording is $0.7875 secs$. Parameters for which data is well approximated with simple fluctuations were obtained by optimization of (\ref{eq:H2}). In order to define $d$, a sample domain is built keeping all parameters fixed and taking $n=1001$ values in $\tau \in [0,4]$.(a) Error - entropy distribution and linear fit ($m=-0.45 , b=-0.96 $). Normal distance to the fit is color coded. (b) Top row: error as a function of $\tau$. Center row: entropy as a function of $\tau$. Bottom row: distance to the linear correlation fitted in (a) as a function of $\tau$. Some local minima have been highlighted for further inspection. (c) Starred solutions. In all panels, green curves correspond to data, blue curves correspond to model output $\{x_i\}$ and red curves correspond to the fluctuations $\{\gamma_i\} = \Omega[\{x_i\}]$. First row: Data (green) and fourier transform calculated separately for each halve of the time domain (indicated b y dashed line). Second row: Model output $\{x_i\} (blue)$ and FFT. Bottom row: Parametric fluctuations and FFT. In both cases, the second half of the dataset is approximated by a subharmonic solution. (d) Circled solutions. Red: simple forcing, bad approximation. Purple: complex forcing, good approximation. Yellow: simple forcing, bad approximation.}
\end{figure*}

As before, we want to show that the parameters so obtained can be found as the minima of a function that implements a parsimony rule. For this we explore a neighborhood of the solution we found by optimizing $H$ taking $n=1001$ equally spaced samples in the domain $\tau = (0,4)$. Although the parameters were found by setting $R=10$ for computational reasons, in order to construct $d$ a higher resolution $R=200$ was used, so that both examples share the same synchronization scheme. Results of this section are summarized in Figure 5. The error of the approximation and the entropy of the fluctuations were calculated at each point in the sample domain. The error-entropy distribution is shown in Fig. 5(a) along with the estimated linear correlation. The corresponding plots of error, entropy and distance to the fit are shown in Figure 5(b). Local minima of $d$ were highlighted with filled circles and filled stars. Figure  5(c,d) shows the model approximation and the fluctuations at each of these locations along with the comparison with data. Green curves correspond to data, blue curves correspond to model output and the red curves are the parametric fluctuations. In order to interpret the results, a Fourier analysis was performed on each time series in (c): this analysis was done independently for each half (dashed line) of the temporal domain so that changes in the signals can be better visualized. Note that in both solutions, the forcing signal fundamental frequency is located to the right of the fundamental frequency of the model output. This suggests that the model might be responding subharmonically to the forcing signal. While both of the starred solutions seem to be interesting, solutions corresponding to the circled local minima fail for different reasons. These are plotted in Figure 5(d) for illustration: the red and yellow ones fail because the approximation is worse than in other points, even though that the fluctuations are simple. The purple one fails because of the excessive complexity of the required fluctuations. Despite that this solution is simpler than its neighbors and also approximates the data, it is more complex than the starred solution which does an equally good job at approximating the data.        

%\begin{widetext} 
%\end{widetext}

It should be clear that the autonomous system (\ref{eq:wc}) will not be able to reproduce data very efficiently: simple inspection of the data indicates that the pressure patterns come in at least two flavors, with and without the wiggle. Since the most similar solution that can be expected from (\ref{eq:wc}) is a limit cycle, we would only be able to adjust the frequency of the cycle in order to minimize the error and both gestures would be approximated by the same morphology. If we insist on explaining the wiggles, better approximations are needed and therefore model (\ref{eq:wc}) should be discarded. One however, could have other reasons to believe that the model is correct and impose the model on the data: this was done in the last section by greatly increasing the model degrees of freedom while still retaining the basic underlying structure.  

In this example, the air sac pressure of a singing bird is approximated by a simple nonlinear system which is being driven by an external signal. We defined an objective function which relies on a mapping between approximations and fluctuations. This allowed us to find model parameters such that the required fluctuations exhibited minimal complexity as measured by $K$. Once this solution was identified, we constructed a parsimony criteria by approximating the scaling relationship between error and entropy. This provided a way to leverage these two quantities and seek for an optimal solution. This balance is measured by function (\ref{eq:H1}) and careful inspection shows that there is not a single `sweet' spot, but many local minima. This means that there are several ways the data can be approximated by simple fluctuations. 

Two mechanisms were found which are interesting and seem to differ from each other. In the first case, the data is approximated at first, as a period (1:1) response to a low frequency forcing. Then, as the wiggle becomes apparent, the model solution resembles a period (1:2) solution. The observation that canary respiratory gestures can be approximated as subharmonic solutions of a driven nonlinear substrate was proposed in \cite{prlsubarmonicos} and later quantified in closely related models \cite{alonso09,alonso10}. There, harmonic input signals were explicitly assumed and parameters were allowed to change in a step-wise manner: it was shown that this and similar models posses subharmonic solutions which yield quantitatively good fits to the data, but the criteria for the choice of the parameters was qualitative. Here, we have made a more generic assumption: that the model is being forced by a specific parameter dependent family of fluctuations. Strikingly, the subharmonic transition was identified as being one of the simplest mechanisms for approximating the data, which is the reason why it corresponds to a local minima of $d$. Moreover, by comparing the balance of error and entropy for neighboring solutions, the choice of the model parameters is justified quantitatively. 

The second mechanism is qualitatively different. At first, the model response is periodic with a period which is similar to period of the forcing. During this epoch, the forcing is also approximately periodic but it is not harmonic. When the pressure pattern changes the model solution is also  very similar to a subharmonic solution but the way the bifurcation occurs is very different. A possible interpretation of this mechanism could read as follows: the forcing signal can be thought to be due to an incoming signal of frequency $f$ plus a copy of the same signal lagged by a phase $\phi$. This suggests that the transition between regimes could be controlled by small changes in the lag. 

This analysis allows us to conclude the following: there are parameters of system (\ref{eq:wc}) for which the solutions of the model explain detailed features of the data at the cost of assuming specific realizations of the forcing signal. Given a search space, we found parameters for which the forcing signal takes the simplest possible form as measured by $K(4)$, within a family of signals which facilitate synchronization with data. By seeking a balance between error and parsimony, two alternative mechanisms were identified. In both cases, the second regime is approximated by a subharmonic solution, but they differ in the way the transition occurs. The possibility that the control signal is changing its frequency was explored quantitatively while careful exploration of the second mechanism remains to be done. Both possibilities are appealing because, within this frameset, they are maximally parsimonious and also they offer a testable prediction: modification of the delicate timing between the substrate response and the forcing signal should lead to very different output patterns \cite{goldin}. 

\section{Concluding remarks}

Estimation of model parameters from experimental time series is a central problem in many disciplines. Our limited ability to perform this task in the case the model is nonlinear can be traced back to the fact that there are no known general procedures for global nonlinear optimization. Furthermore, in an experimental setting we are likely to be uncertain  about the functional form of the underlying dynamical rules. Moreover, in the case of physiological data, it is likely that the system under observation is taking input from external sources so it cannot be considered in isolation. Therefore, investigating parametric fluctuations can be relevant in some situations of interest. 

In this work we proposed a nonlinear transformation for time series and motivated its usefulness. We draw inspiration from an ideally continuous case to define the transformation $\Omega$ of a given dataset. This transformation takes the observed time series and returns the parametric fluctuations that yield solutions which optimally match the time series. It is clear that in many situations better approximations can be achieved by allowing arbitrary fluctuations in the parameters: this follows from the fact that considering arbitrary fluctuations yields a system with infinite degrees of freedom which can accommodate any dataset. However, due to nonlinearity, this transformation is highly dependent on the functional form and parameter values of model $F(x,p)$. Here we showed in a case study, that for specific nonlinearities coded in $F(x,p)$ the fluctuations take a simple form. This transformation was applied to the problem of parameter estimation in a case study with both numerical and experimental data. 

We tailored  our procedure to address the question of whether a system is making use of nonlinear mechanisms to code for different functionalities. The idea that controlling the unstable periodic orbits in chaotic attractor could serve as a strategy to generate qualitatively different behavior was proposed by Ott et. al.\cite{ottcontrol} and demonstrated experimentally in a chemical system \cite{showalter}. Recent work in the field of birdsong suggests that much of the complexity observed in the song and respiratory patterns of canaries can be successfully explained by considering simple time dependent unfoldings of low dimensional nonlinear systems \cite{simplemotorgestures,prlsubarmonicos,alonso09,alonso10,yonizeke2011}. 

This ideas were tested in a simple rate coding model for populations of neurons. Data was generated by driving the model with simple harmonic signals. Small changes in the forcing signal lead to qualitatively different behavior: this is the scenario in which these methods seem to be useful. Then, we proceeded to ask if the general structure of the unperturbed model (\ref{eq:wc}) held any relationship with the numeric data. By construction, solutions of the autonomous model fail to approximate the datasets. However, considering the alternative hypothesis that fluctuations were present at the time of the measurements, we found that there is a range in parameter space for which the cost of introducing this hypothesis is minimal. These ranges correspond roughly to the values of the parameters used to generate the dataset. Therefore we conclude that local minima of $d(x_0,p)$ potentially provide insight about the underlying dynamics which generated the data. We have also studied a closely related function which does not require evaluation in a sample domain and yields the same answer for a range in the threshold values. This function was used as an objective function for model identification when we considered experimental data. Model (\ref{eq:wcgenforcing}) was trained to satisfy a threshold condition for the quality of the approximations by using simple fluctuations. We found ranges in parameter space for which the scaling relationship between error and entropy breaks maximally. Inspection of these minima revealed two qualitatively different solutions. Interestingly, in both cases data is approximated by a solution which resembles very much a bifurcating period 2 solution: the difference lies in the way this bifurcation is controlled.

The fact that data could be approximated by a forced low dimensional system is very interesting from the theoretical point of view. On the one hand, if the systems are low dimensional, there are methods to classify families of models according to the geometrical mechanisms which underlie the generation of complex behavior \cite{classification, alicestretchland}. In particular, period doubling solutions have been found experimentally in many physical systems and they can be associated to universal scaling laws \cite{feigenbaum,libchaber}. On the other hand, it has been shown that even in the case of having many nonlinear interacting units, the resulting macroscopic average behavior can follow low dimensional dynamical rules. A recent breakthrough in the analysis of Kuramoto's model due to Ott and Antonsen allows to obtain mean field equations for classes of closely related models \cite{ott2008,strogatz2008} . In some situations, the mean field equations exhibit non trivial bifurcation diagrams which share many features with system (\ref{eq:wc}). It has been shown that there are conditions under which the average behavior of a driven set of excitable units displays complex behavior by dynamical variations of the degree of synchrony in the driven population \cite{alonso11}. More recently, it was shown that by using a mean field description, algorithms can be designed that are successful at controlling these attractors \cite{so2014}. Thus, there is hope that much of the complexity in macroscopic phenomena can be captured by low dimensional nonlinear mechanisms. The purpose of this work is to further bridge the gap between the description and characterization of qualitative nonlinear mechanisms and quantification. Finally, we note that the same mapping can be used to explore how a given nonlinear system may optimize other quantities of interest by considering alternative metrics on the parametric fluctuations. 

\section{Acknowledgements}
This work was funded by NSF grant EF-0928723 awarded to Marcelo O. Magnasco. The author acknowledges the support and mentorship of Gabriel B. Mindlin and Marcelo O. Magnasco, and both review and helpful discussions with Carl D. Modes. Data was kindly provided by the Dynamical systems lab at University of Buenos Aires. 

\appendix
\section{General implementation}

Here we describe a numerical procedure to build approximations to the problem of finding minima of the high dimensional error function (\ref{eq:highdimcost}) defined in section 3. The strategy is to compute $\Omega$ in small running windows so that brute force optimization is possible. Here we state a more general definition of function (\ref{eq:highdimcost}) which allows better allocation of computational effort. 

Before we allowed the fluctuations to take any of $R$ values in a domain at each time step. Now, we allow the fluctuations to take $R$ values in $b$ time bins where $b$ is an integer such that $N \mod b = 0$. 

\begin{equation}
\label{eq:general}
E_{x_0, p} (\{\gamma_0 . . .  \gamma_b \} ) = \frac{1}{N} \sum_{i=0}^{N} D(\phi_{t_i}(x_0,p+\gamma_{\floor*{\frac{b*i}{N}}  }), o_i)
\end{equation}

Let $\{\gamma_0 . . .  \gamma_b\}$ be the minimum of (\ref{eq:general}). We define the transformation $\Omega$ as 

\begin{equation}
\label{eq:generalomega}
\Omega[\{o_i\}]_i = \gamma_{\floor*{\frac{b*i}{N}} } 
\end{equation}

And the corresponding model output

\begin{equation}
\label{eq:inverseomega}
\Omega^{-1}_{x_0,p} (\{\gamma_i\}) =\vcentcolon \{ \phi_{t_i}(x_0,p+\gamma_{\floor*{\frac{b*i}{N}}} ) \} = \{ x_i \}
\end{equation}

Summarizing, we apply definitions (\ref{eq:general}), (\ref{eq:generalomega}) and (\ref{eq:inverseomega}) in small running windows of size $d \ll N$. The window is advanced $s$ steps and a new $\Omega$ is determined. By concatenating the results of this procedure we construct approximations to the minima of (\ref{eq:general}). In this notation, results of this articles were obtained by setting $R=200$, $d=2$, $s=1$, $b=2$. This procedure is illustrated in Figure 6. 

%\begin{widetext} 
\begin{figure*}[hbt!]
\label{fig:appendix}
\includegraphics[width=100mm]{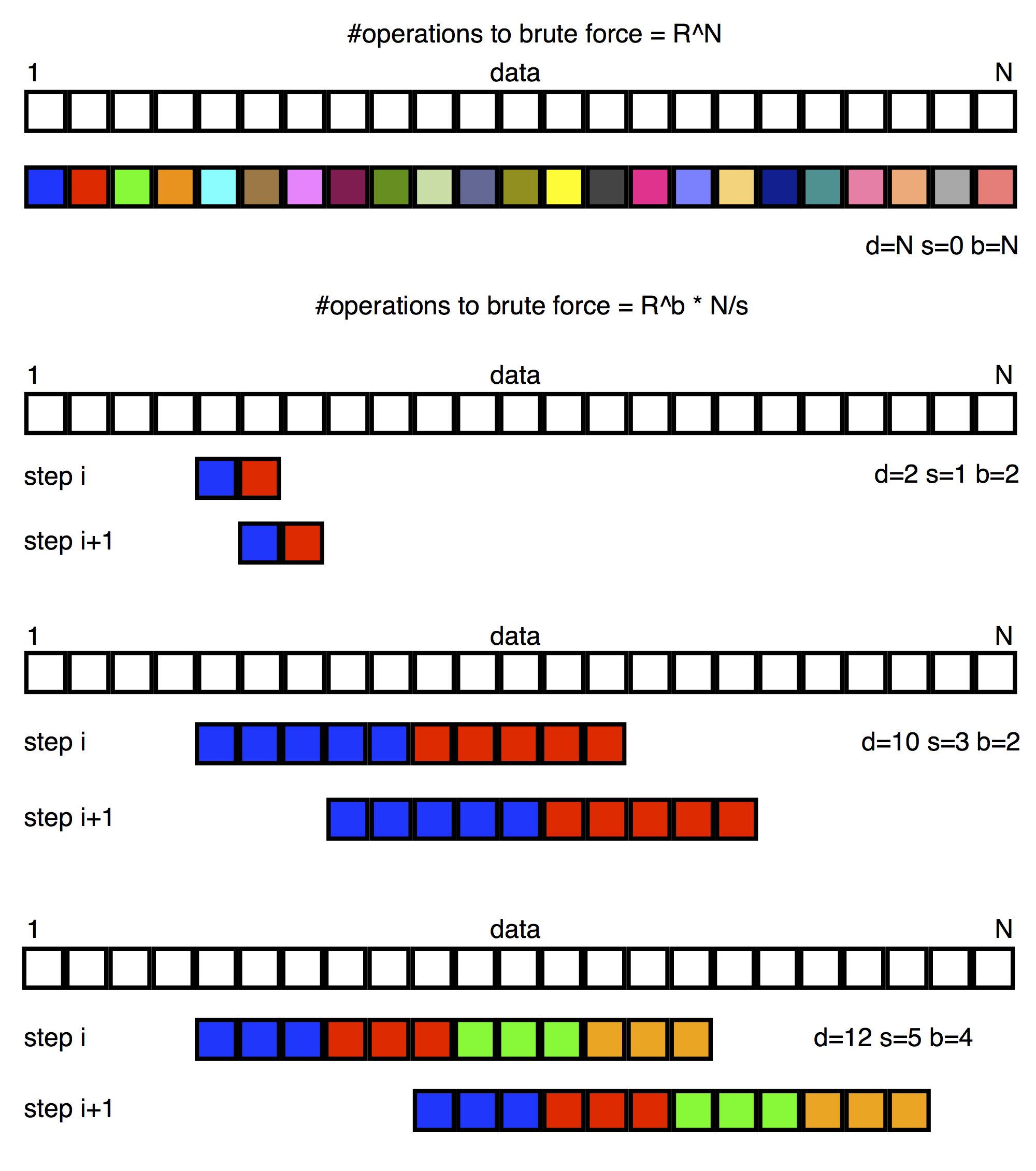}
\caption{Numerical implementation. Data is represented by the empty squares. (a) Arguments of function (\ref{eq:general}) are represented by colored squares. For each color, the fluctuations can take $R$ equally spaced values in the search domain specified in $\epsilon$. Attempts to optimize this function by brute force fail for large $N$. Thus, approximations to (\ref{eq:generalomega}) are built by applying defs. (\ref{eq:general}),(\ref{eq:generalomega}) and (\ref{eq:inverseomega}) in small running windows with $d \ll N$. (b) At each window, we allow the fluctuations to take $R$ independent values in $b$ bins, where  $b$ is such that $d \mod b = 0$. The entries to function (\ref{eq:general}) are represented by colored blocks. As before, different colors indicate independent values for the fluctuations. For different choices of $b$, we reduce the number of independent values for $\gamma_i$ and therefore larger values of $d$ become computationally treatable. The approximation is built by advancing the window $s$ steps and concatenating the outputs.}
\end{figure*}
%\end{widetext}

\end{document}